\documentclass[a4paper,11pt]{article}
\pdfoutput=1 % if your are submitting a pdflatex (i.e. if you have
\usepackage{jcappub} % for details on the use of the package, please
\usepackage[T1]{fontenc} % if needed

\usepackage{diagbox}
\usepackage{multirow}
\usepackage{amsmath}
\usepackage{amssymb}

\newcommand{\de}{\mathrm{d}}
\renewcommand{\Vec}[1]{\mathbf{#1}}

\newcommand{\Min}{\textsc{Minerva}}
\newcommand{\Pin}{\textsc{Pinocchio}}

\newcommand{\hMpc}{h^{-1} \,\mathrm{Mpc}}
\newcommand{\kMpc}{\, h \, {\rm Mpc}^{-1}}
\newcommand{\cGpc}{\, h^{-3} \, {\rm Gpc}^3}

\usepackage{color}

\usepackage[normalem]{ulem}

\title{\boldmath The halo 3-point correlation function: a methodological analysis.}

\author[a,1]{A. Veropalumbo,\note{Corresponding author.}}
\author{A. Binetti,}%[b]
\author[c,d]{E. Branchini,}
\author[e,f]{M. Moresco,}
\author[g,h,i,l]{P. Monaco,}
\author{A. Oddo,}%[h,l]
\author[k]{A. G. S\'anchez,}
\author[h,i,j]{E. Sefusatti}

\affiliation[a]{Dipartimento di Fisica, Universit\`a di Roma Tre, Via della Vasca Navale 84, I-00146 Roma, Italy}
\affiliation[b]{Dipartimento di Fisica "Aldo Pontremoli", Universit\`a degli Studi di
Milano, Via Celoria 16, I-20133 Milano, Italy}
\affiliation[c]{INFN - Sezione di Roma Tre, Via della Vasca Navale 84, I-00146 Roma, Italy}
\affiliation[d]{Dipartimento di Fisica, Universit\`a degli Studi di Genova, and INFN Sezione di Genova, Via Dodecaneso 33, I-16146, Genova, Italy}
\affiliation[e]{Dipartimento di Fisica e Astronomia "Augusto Righi" - Alma Mater
Studiorum Universit\`a di Bologna, Via Piero Gobetti 93/2, I-40129
Bologna, Italy}
\affiliation[f]{ INAF-Osservatorio di Astrofisica e Scienza dello Spazio di Bologna,
Via Piero Gobetti 93/3, I-40129 Bologna, Italy}
\affiliation[g]{Dipartimento di Fisica - Sezione di Astronomia, Universit\`a di Trieste,
Via Tiepolo 11, I-34131 Trieste, Italy}
\affiliation[h]{
IFPU, Institute for Fundamental Physics of the Universe, Via Beirut 2,
I-34151 Trieste, Italy}
\affiliation[i]{
INAF-Osservatorio Astronomico di Trieste, Via G.B. Tiepolo 11, I-34143 Trieste, Italy}
\affiliation[j]{INFN, Sezione di Trieste, Via Valerio 2, I-34127 Trieste, Italy}
\affiliation[k]{Max Planck Institute for Extraterrestrial Physics, Giessenbachstr. 1,
D-85748 Garching, Germany}
\affiliation[l]{SISSA - International School for Advanced Studies, Via Bonomea 265, I-34136 Trieste, Italy}

\emailAdd{alfonso.veropalumbo@uniroma3.it}

\abstract{
 Upcoming galaxy surveys will provide us with an unprecedented 
view of the Large-Scale Structure of the Universe and the realistic chance to extract valuable astrophysical and cosmological information from higher-order clustering statistics. 
This perspective poses new challenges, requiring both accurate and efficient estimators and a renewed assessment of possible systematic errors in the theoretical models and likelihood assumptions. This work investigates these issues in relation to the analysis of the 3-point correlation function (3PCF) in configuration space.
We measure the 3PCF of 300 halo catalogs from the {\Min} simulations covering a total volume of $~1000 h^{-3} \mathrm{Gpc}^3$. Each 3PCF measurement includes {\em all} possible triangular configurations with sides between 20 and $130\hMpc$. In the first place, we test different estimates of the covariance matrix, a crucial aspect of the analysis. We compare the covariance computed numerically from the limited but accurate benchmark simulations set to the one obtained from $10000$ approximate halo catalogs generated with the {\Pin} code. We demonstrate that the two numerically-estimated covariance matrices largely match, confirming the validity of approximate methods based on Lagrangian Perturbation Theory for generating mocks suitable for covariance estimation. We also compare the numerical covariance with a theoretical prediction in the Gaussian approximation. We find a good match between the two for separations above 40$\hMpc$. 
We test the 3PCF tree-level model in Perturbation Theory. The model is adopted in a likelihood analysis aimed at the determination of bias parameters. 
We find that, for our sample of halos at redshift $z=1$, the tree-level model performs well for separations $r \geq 40 \hMpc$. Results obtained with this scale cut are robust against different choices of covariance matrix. We compare to the analogous analysis of the halo bispectrum already presented in a previous publication, finding a remarkable agreement between the two statistics. We notice that such comparison relies, to the best of our knowledge for the first time, on a robust and consistent covariance estimate and on the inclusion of essentially all measurable configurations in Fourier as in configuration space.
We then test different assumptions to build the model defining a robust combination of hypotheses that lead to unbiased parameter estimates. 
Our results confirm the importance of 3PCF, supplying a solid recipe for its inclusion in likelihood analyses. Moreover, it opens the path for further improvements, especially in modelling, to extract information from non-linear regimes.
}

\begin{document}
\maketitle
\flushbottom
%==================================================
\section{Introduction}
\label{sec:intro}
It has long been realized that the analysis of the large-scale structure (LSS) of the Universe traced by the
spatial distribution of galaxies provides a wealth of information complementary to other cosmological probes as weak gravitational lensing or the Cosmic Microwave Background. As a result, several galaxy redshift surveys have been carried out and planned to probe an increasingly larger fraction of the observable Universe, mapping the spatial distributions of tens of millions of galaxies  (e.g. DESI \cite{DESI1, DESI2}, the Euclid spectroscopic survey \cite{Euclid} and the Roman Telescope high-latitude survey \cite{Roman}). So far, measurements of the of two-point statistics have been the main tool for the analysis of the LSS. Such statistics can be either the power spectrum in Fourier space, or the 2-point correlation function in configuration space. Yet, larger survey volumes will allow more precise measurements of higher-order clustering statistics, able to quantify the non-Gaussian nature of cosmological perturbations at low redshift. Such measurements can therefore help to disentangle different sources of non-Gaussianity, either of intrinsic interest if it's of primordial origin or possibly responsible for relevant degeneracies with cosmological parameters in the case of non-linear galaxy bias.

Indeed, the analysis of higher-order statistics in recent datasets is becoming routine, particularly in the case of the galaxy bispectrum, the 3-point correlation function in Fourier space \cite{GilMarin2017, DAmicoEtal2020, DAmicoEtal2022A, PhilcoxIvanov2022, CabassEtal2022A, CabassEtal2022B}. The bispectrum, in fact, presents some advantages with respect to its configuration-space counterpart. In the first place, the computational cost of measuring the bispectrum of the largest galaxy catalogs is moderate, allowing one to accurately compute its covariance matrix if an adequate number of simulated catalog is available \cite{SefusattiEtal2006, ChanBlot2017, GilMarin2017, Colavincenzo2019, HahnEtal2020, Oddo2020, GualdiGilMarinVerde2021}. In the second place, theoretical models for the bispectrum, in some cases even up to one loop in Perturbation Theory, have been extensively validated with large, simulated data-sets \cite{Oddo2020, Oddo2021, BaldaufEtal2021, MoradinezhadDizgahEtal2021, IvanovEtal2021A, Alkhanishvili2021, EggemeierEtal2021, RizzoEtal2022A}. 

Similar developments have taken place for three-point correlation function (3PCF) in configuration space at a  relatively slower pace due to the higher computational cost of its estimation in large galaxy catalogs and simulations (\cite{Jing1997, Barriga2002, Bel2015, Hoffmann2018, Kuruvilla2020} and references therein). In this respect, the situation has changed dramatically with the advent of a new class of efficient 3PCF estimators \citep{Slepian2015, Slepian2018, Sugiyama2019, Umeh2021} that reduced the computational scaling cost from $N^3$ to $N^2$, where $N$ is the number of objects in the catalog. This  allowed the analysis of large datasets and for the first detection of baryonic acoustic oscillation features in the 3PCF of galaxies \citep{Slepian2017a} and clusters \cite{Moresco2021}, and to perform a joint two- and three-point correlation analysis \citep{Veropalumbo2021} paralleling those performed in Fourier space \cite{GilMarin2017}. Such advances triggered a development on the theoretical side that pushed the 3PCF model beyond the analysis of the sole monopole, taking advantage of its redshift-space anisotropy \citep{Slepian2017a, Slepian2017, Sugiyama2021}. 

The main advantage of the 3PCF over the bispectrum is its ability to deal with survey geometry. The footprint of spectroscopic surveys is characterized by irregular boundaries and holes in correspondence to regions that could not be observed. A complex survey geometry can be naturally dealt with in configuration space by using appropriate estimators as, in the 3PCF case, the Szapudi-Szalay one \cite{Szapudi1998}. This is not the case in Fourier space, where the survey footprint induces mode coupling that needs to be accounted for by either de-convolving the measured correlation signal or by convolving the theory prediction with the window \cite{Sugiyama2019, Philcox2021b, PardedeEtal2022A}.
 
All these considerations justify the renewed interest in the 3PCF statistics and the urgency to carry out a dedicated study aimed at investigating how various methodological choices in the 3PCF analysis can affect the inference of model parameters.
In this work we will assess the effect of these choices on the determination of galaxy bias parameters, one of the traditional motivations to study the 3PCF.

The most relevant one is the estimation of the covariance matrix. Numerical estimates of the 3PCF covariance matrix that rely on individual measurements from a set of mock catalogs are time-consuming, both because of mocks production and 3PCF measurements, despite the fast estimator mentioned above. 
A number of workarounds have been proposed to reduce the computational cost while keeping the accuracy at an acceptable level. Computational efficiency can be increased either by speeding up the creation of mocks by means of approximate halo catalogs (see \cite{Monaco2016} for a recent review) or by approximating the covariance matrix itself.
In this work we will explore both possibilities.
First of all, we will compare the 3PCF covariance matrix estimated from catalogs of dark matter halos obtained from the {\Min} set of full N-body simulations to the one estimated from a set of halo catalogs obtained with the approximated method implemented in the {\Pin} code, based on Lagrangian Perturbation Theory \cite{Monaco2002, MunariEtal2017}.
Then we shall compare the performances of different 3PCF covariance estimators,
considering the brute-force numerical estimate from mocks, the analytical Gaussian model of \cite{Slepian2015} and the shrinkage method \citep{Pope2008}, i.e. a hybrid combination of numerical and analytical estimates. To evaluate the goodness of each method we will perform 
the same likelihood analysis with different covariance matrices and compare the recovered best-fit estimates of the halo bias parameters.
We will explore as well additional aspects of the standard 3PCF analysis. One is the selection of the triangular configurations included in the analysis. A related one is the choice of the separation bin size, defining such triangles. In both cases we will evaluate the impact on the determination of constraints of the bias parameters. Finally, we will consider different relations among bias parameters, either theoretically-motivated or based on fits to numerical simulations, as a way to reduce the parameter space. 

Our results will ultimately provide an extensive assessment of the validity of the tree-level PT model of the 3PCF under a variety of assumptions. The posterior constraints on bias parameters will be compared to the similar results obtained in \cite{Oddo2020} from the halo bispectrum measured from the same data set. This constitutes, to the best of our knowledge, the most direct and fair comparison between the two statistics presented so far in the literature, based, {\em on both sides}, on all measurable configurations, as a function of the smallest scale included, and taking advantage of a robust estimate of the covariance properties from the same set of approximate mocks.

The  paper is structured as follows: in Sec.~\ref{sec:data} we briefly 
describe the two halo catalogs used in our analysis and their parent simulations. In Section~\ref{sec:meas} we describe the estimators used to measure the 3PCF and its covariance matrix, then we compare the covariance matrices numerically estimated from the two different sets of mock catalogs obtained from different types of simulations.
In Sec.~\ref{sec:model} we present the model for the halo 3PCF adopted in this work and used, along with the measured 3PCF and its covariance matrix, to perform the likelihood analysis described in Sec.~\ref{sec:paraminf} and to estimate the halo bias parameters.
The results of this analysis, aimed at assessing the impact of the various methodological options described above are presented in Sec.~\ref{sec:results} and discussed in Sec.~\ref{sec:concl}, in which we draw our main conclusions. 

%==================================================
\section{Data sets}
\label{sec:data}

All results presented in this work are based on two different sets of halo catalogs. The first is constituted by 298 catalogs of dark matter halos obtained from the {\Min} set of full N-body simulations. The second is a much larger set of 10,000 approximate halo catalogs produced with the {\Pin} code. Both sets, sharing the same simulation box and cosmological parameters, are described below.  

%==================================================
\subsection{{\Min} halo catalogs}
\label{sec:minerva_sim}

The {\Min} simulations \cite{Grieb2016, Lippich2019} have been performed with the \textsc{GADGET-III} code \cite{Springel2005} evolving $1000^3$ dark matter particles of mass $m_p \simeq 2.67 \cdot 10^{10} \, h^{-1} \mathrm{M}_{\odot}$ in a cubic box of side $L=1500 \, \hMpc$.

Dark matter halos were identified with a Friends-of-Friends algorithm of linking length equal to $0.2$ times the mean inter-particle distance.
Following \cite{Oddo2020}, we only consider the $z=1$ output and halos with masses larger than $1.12 \cdot 10^{13} \, h^{-1} \mathrm{M}_{\odot}$, corresponding to a mean number density ($\Bar{n} = 2.13 \cdot 10^{-4} \, h^{3} \mathrm{Mpc}^{-3}$), roughly matching that of H-$\alpha$ line emission galaxies that will be observed in wide surveys like Euclid \cite{Euclid}.

%==================================================
\subsection{{\Pin} halo catalogs}
\label{sec:pinocchio_mocks}

The second set of halo catalogs has been obtained with the {\Pin} code \cite{Monaco2002, MunariEtal2017}. Unlike the {\Min} set, here the dynamical evolution of the cosmic density field is approximated using third order Lagrangian Perturbation Theory. Dark matter particles are grouped in halos using a procedure based on the ellipsoidal collapse model. Since {\Pin} realisations are computationally cheaper than N-body simulations, a larger number of mock halo catalogs can be generated.  Here we consider the same $10,000$ catalogs used by \cite{Oddo2020} and adopt the same threshold mass, originally chosen to match the amplitude of the halo power spectrum measured in the {\Min} simulations at large scales. It has been shown that this matching minimizes the difference between the power spectrum and bispectrum variance estimated from the {\Pin} mocks and the one from the {\Min} simulations, both in real \cite{Oddo2020, Oddo2021} as in redshift-space \cite{RizzoEtal2022A}.
We highlight that the first 300 {\Pin} mocks share the same random seed for the initial condition of the {\Min} simulations. This allows a direct comparison that is not affected by sample variance.

%==================================================
\section{Measuring the 3PCF}
\label{sec:meas}
%==================================================

\subsection{3PCF estimator}
\label{sec:zetameas}

To measure the 3PCF of the DM halos in the {\Pin} and {\Min} catalogs we use the Szapudi \& Szalay estimator 
\cite{Szapudi1998} as implemented in \cite{Slepian2015}. That estimator relies on the local Spherical Harmonics Decomposition (SHD) of the 3PCF
\begin{equation}
    \label{eq:zeta_ell}
    \zeta (r_{12}, r_{13}, \mu) \, = \sum_{\ell}^{\ell_{max}} \zeta_{\ell} (r_{12}, r_{13}) \,\mathcal{L}_{\ell}(\mu) \, ,
\end{equation}
where $r_{12}$ and  $r_{13}$ are two sides of a triangle while $\mu$ is the cosine of the angle between them. The length of the third side is set by the vector relation $\Vec{r}_{12}+\Vec{r}_{13}+\Vec{r}_{23}=0$.
$\mathcal{L}_{\ell}(\mu)$ is the $\ell$-th Legendre polynomial.
The SHD algorithm measures the coefficients of the expansion
 $\zeta_{\ell}(r_{12}, r_{13})$ up to a maximum multipole $\ell_{max}$. Therefore eq.~\ref{eq:zeta_ell} provides an approximated estimate of the halo 3PCF. However, for a typical galaxy catalog, $\ell_{max}=10$ is sufficient to converge to the correct solution \citep{Veropalumbo2021}.
 
For the multipoles measurements $\hat{\zeta}_\ell(r_{12}, r_{23})$, we consider linear bins of size $\Delta r = 10\, \hMpc$ up to a maximum separation   $r_{max} = 150 \,\hMpc$.
 The standard 3PCF is then obtained as 
 \begin{equation}
    \label{eq:zeta_ell_tr}
    \hat{\zeta} (r_{12}, r_{13}; r_{23}) \, = \sum_{\ell}^{\ell_{max}} \hat{\zeta}_{\ell} (r_{12}, r_{13}) \,\widetilde{\mathcal{L}}_{\ell}(r_{12}, r_{13}, r_{23}) \, ,
\end{equation}
where $\widetilde{\mathcal{L}}_{\ell}(r_{12}, r_{13}, r_{23})$ is the Legendre polynomial of order $\ell$ weighted over the triangle
\begin{equation}
    \label{eq:leg_ave}
        \widetilde{\mathcal{L}}_{\ell}(r_{12}, r_{13}, r_{23}) = \frac{
        \displaystyle\int_{r_{12} - \Delta_r/2}^{r_{12} +  \Delta_r/2} \mathrm{d} p p^2
        \displaystyle\int_{r_{13} - \Delta_r/2}^{r_{13} +  \Delta_r/2} \mathrm{d} q q^2
        \displaystyle\int_{r_{23} - \Delta_r/2}^{r_{23} +  \Delta_r/2} \mathrm{d} s s^2
        \Theta(p, q, s) \mathcal{L}_{\ell}(p, q, s)}
        {\displaystyle\int_{r_{12} - \Delta_r/2}^{r_{12} +  \Delta_r/2} \mathrm{d} p p^2
         \displaystyle\int_{r_{13} - \Delta_r/2}^{r_{13} +  \Delta_r/2} \mathrm{d} q q^2
         \displaystyle\int_{r_{23} - \Delta_r/2}^{r_{23} +  \Delta_r/2} \mathrm{d} s s^2
         \Theta(p,q,s)}
\end{equation}
with 
\begin{equation}
    \label{eq:theta}
    \Theta(r_{12}, r_{13}, r_{23}) = 
\left\{ 
  \begin{array}{ c l }
    1 & \quad \textrm{if} \, \left| \frac{r_{12}^2 + r_{13}^2 - r_{23}^2}{2 r_{12} r_{13}} \right| \leq 1 \\
    0 & \quad \textrm{otherwise}
  \end{array}
\right.
\end{equation}
and it accounts for the binning of the third side \citep{Veropalumbo2021}. The estimator cannot properly recover the 3PCF corresponding to isosceles configurations ($r_{12}=r_{13}$) with a finite and manageable number of multipoles. In fact, this intrinsic limitation extends to nearly isosceles triangles where $r_{13}-r_{12}\simeq \Delta r$. For this reason, we characterise a given set of measurements not only in terms of the minimum separation $r_{min}$, but also by a minimum value $\eta_{min}$ for the relative difference
\begin{equation}
\label{eq:eta}
    \eta \equiv \frac{r_{13} - r_{12}}{\Delta r}\,.
\end{equation}
In other terms, fixed $\Delta r$, $r_{min}$ and $r_{max}$, setting $\eta_{min}=0$ would correspond to measure all available triangles with sides within the given range in steps of $\Delta r$. $\eta_{min}=1$ would exclude configurations with $r_{13}=r_{12}$, while higher values would further reduce the triangle set. As we will see our ability to properly model the 3PCF measured with the SHD estimator depends significantly on this parameter. In Sec.~\ref{sec:paraminf} we perform several tests to assess the sensitivity of the 3PCF analysis to the choice of both $r_{min} $ and $\eta_{min}$.

Input to the estimator are a catalog of halos and a catalog of random objects, that we created, which are distributed in the same volume but lacking clustering properties. 
We've performed the measurements with the random splitting method. With this technique, we measure the 3PCF with many small random samples and then take their average. We impose the data-random ratio after splitting $N_R=1$. Since the samples have different numbers of objects, we change the random accordingly. In the {\Min} 3PCF measurements the number of random objects is 50 times larger than that of the halos. In the {\Pin} case, for computational constraints, we use 10 times as many objects. For this reason, even for the low density randoms, we do not expect a biased 3PCF estimation but only an additional contribution to its variance. The effect on the total error budget, however, is negligible.
We notice that it's not straightforward to change the estimator to exploit the periodicity of the box, thus avoiding the use of a random sample. First, the Szapudi-Szalay estimator directly measure the connected part of the 3PCF, internally subtracting the disconnected part. Second, since we use \cite{Slepian2015} for the triplet counts, we measure the multipoles of the 3PCF. We've tried to estimate the full 3PCF as $DDD/N_{D}^3$ normalized by the analytic triplet counts, with $DDD$ the data triplets and $N_{D}$ is the total number of triplets. We then need to subtract the disconnected part, that is 1+$\xi(r_{12})+\xi(r_{13})+\xi(r_{23})$. We've verified that this procedure is not sufficiently precise for our setting.

%==================================================
\begin{figure}
    \centering
    \includegraphics[width=0.9\textwidth]{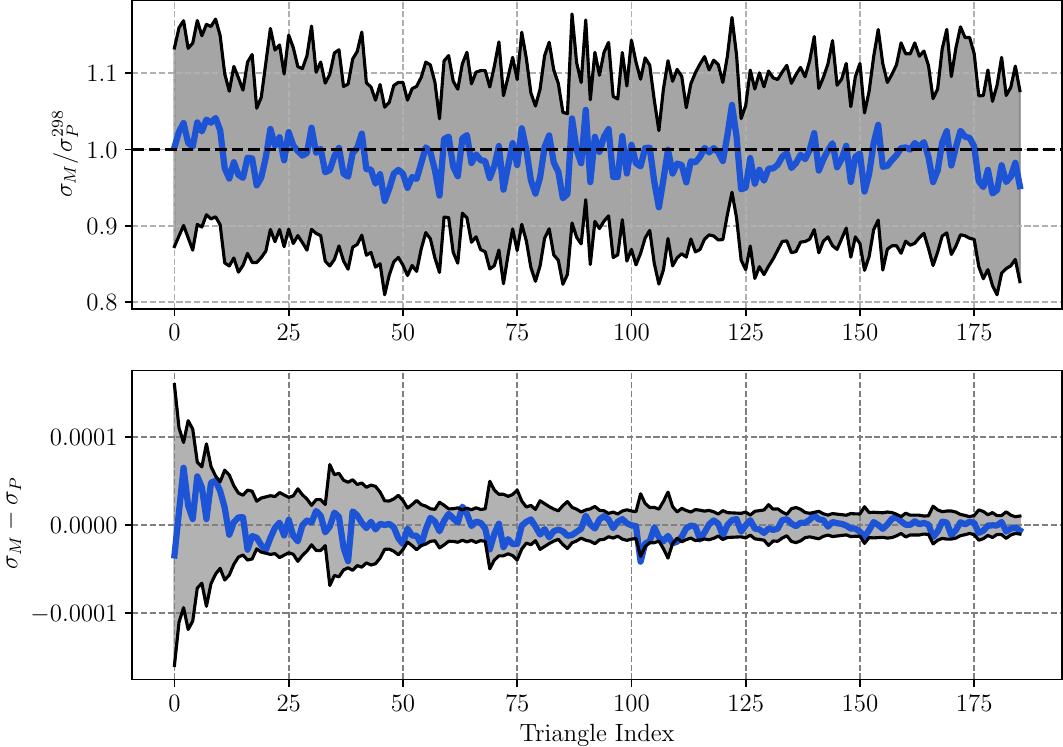}
    \caption{\textit{Top panel}. Blue curve:
    ratio between the 3PCF {\it rms} scatter measured in the {\Min}, $\sigma_M$, and measured in the first 298 {\Pin} mock catalogs, $\sigma_P^{298}$, shown as a function of the triangle type, identified by an index. Larger indexes identify larger triangles. Grey band: $2-\sigma$ scatter of  $\sigma_M/\sigma_P^{298}$, obtained propagating the variance over single {\it rms}. 
    \textit{Bottom panel}. Blue curve: difference between the 3PCF {\it rms} scatter measured the {\Min}, $\sigma_M$, and measured in the {\Pin} mock catalogs, $\sigma_P$, shown as a function of the triangle type, identified by an index. Grey band: the reference $2-\sigma$ scatter was obtained from subsets of 298 realisations extracted from the 10,000 {\Pin} mocks. See text for details.}
    \label{fig:std}
\end{figure}
%==================================================
\begin{figure}
    \centering
    \includegraphics[width=0.8\textwidth]    {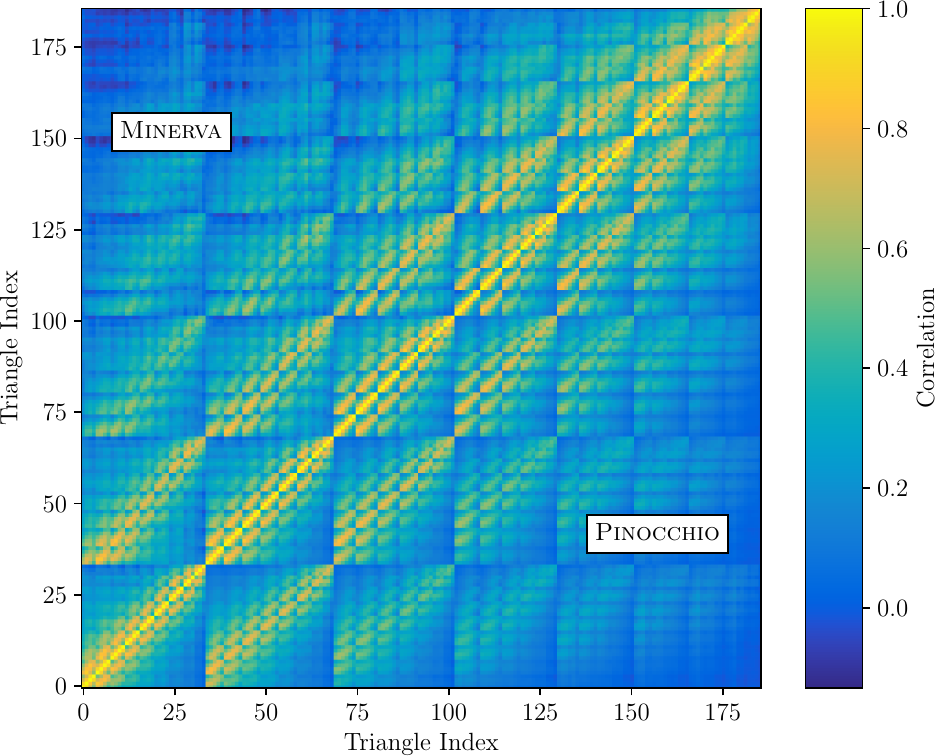}
    \caption{3PCF correlation matrix measured from the {\Min} mock catalogs
      (upper triangular matrix) and from the 10000  {\Pin} mocks 
      (lower triangular matrix). Indexes identify different triangle configurations. Larger indexes indicate larger scales. The correlation amplitude is color-coded as indicated in the color bar.}
    \label{fig:correlation}
\end{figure}
%==================================================
\begin{figure}
    \centering
    \includegraphics[width=0.9\textwidth]{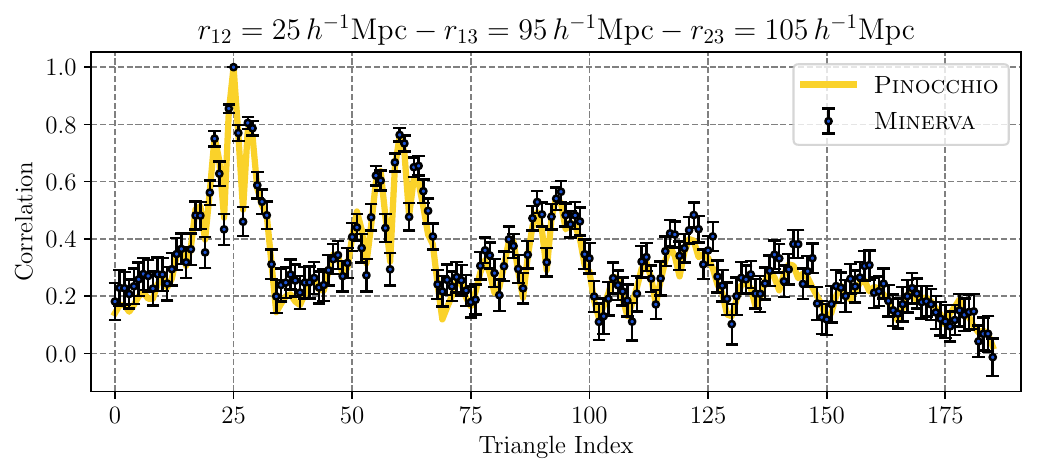} \\
     \includegraphics[width=0.9
     \textwidth]{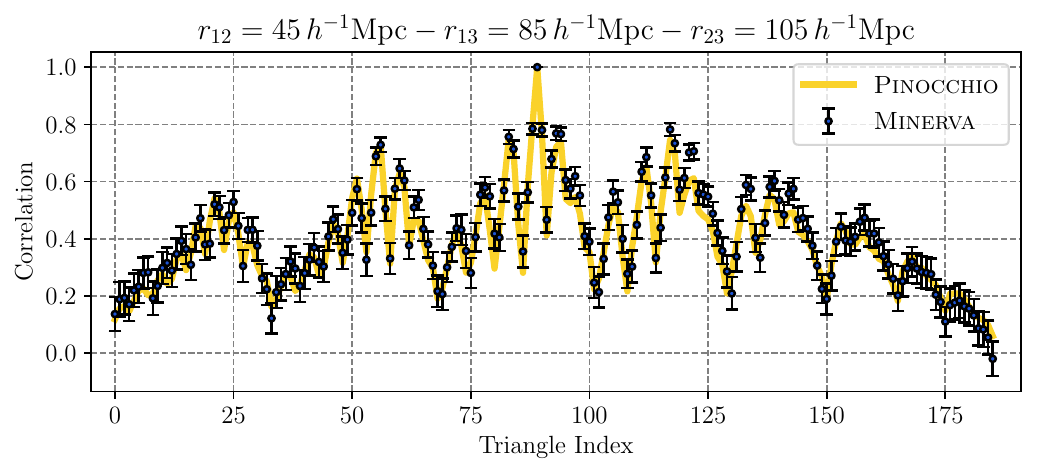} \\
    \caption{\textit{Top panel}: correlation amplitude of the elements along a cut through 
     correlation matrix in correspondence of a triangle with sides $25 \, \hMpc, 95 \, \hMpc, 105 \, \hMpc$. Black dots: {\Min} estimates. Yellow curve: {\Pin} estimates.
     Error bars:  scatter off subsets of 298 realisations extracted from the 10,000 {\Pin} mocks. \textit{Bottom panel}: same quantity estimated for a larger triangle with sides $45 \, \hMpc, 85 \, \hMpc, 105 \, \hMpc$.}
    \label{fig:corr_rows}
\end{figure}

%==================================================

\subsection{Covariance matrix}
\label{sec:meascov}

The numerical covariance matrix is estimated from the 3PCFs measured in each mock catalog as follows:
\begin{equation}
    \label{eq:cov}
    \widehat{C}_{i, j} = \frac{1}{N_{mocks}-1} \sum_{n=1}^{N_{mocks}} \left(\hat{\zeta}_i^n-\overline{\zeta}_i \right) 
    \left(\hat{\zeta}_j^n-\overline{\zeta}_j \right)
\end{equation}
where the indexes $i$ and $j$ identify any two triangles in the sample with sides 
 $\lbrace r_{12}, r_{13}, r_{23} \rbrace$ and $\lbrace r_{12}^{\prime}, r_{13}^{\prime}, r_{23}^{\prime} \rbrace$. 
$\hat{\zeta}^n$ indicates the
3PCF measured in the $n$-th mock and $\overline{\zeta}$ the average among the mocks
\begin{equation}
    \label{eq:mean_zeta}
    \overline{\zeta}_i = \frac{1}{N_{mocks}} \sum_{n=1}^{N_{mocks}} \hat{\zeta}_i^n.
\end{equation}
Then, to compare the {\Min} and {\Pin} estimates we consider as well the correlation matrix, defined as
\begin{equation}
    \label{eq:corr}
    c_{ij} \equiv \frac{\widehat{C}_{ij}}{\sqrt{\sigma_i\sigma_j}}
\end{equation}
where $\widehat{C}_{i,j}$ is the covariance matrix element and $\sigma_i \equiv \sqrt{\widehat{C}_{i,i}} $ represents the dispersion of the measurements for triangle $i$.

\subsection{3PCF covariance:  {\Min} vs {\Pin} }
\label{sec:MinervavsPinocchio}

In this section we provide a first comparison between the numerical covariance estimated from the {\Min} N-body simulations and the one estimated from the {\Pin} approximate mocks. 

The top panel of Fig.~\ref{fig:std} shows the ratio between the {\Min} variance and the variance from the first 298 {\Pin} mocks sharing the same initial conditions. Each measurement corresponds to all triangles with $r_{min} = 20 \hMpc, r_{max} = 130 \hMpc, \eta = 1$ ordered assuming $r_{12} < r_{13} \leq r_{23}$. The shaded area shows the 2-$\sigma$ scatter obtained from subsets of 298 realisations extracted from the 10,000 {\Pin} mocks. The bottom panel shows instead the difference between the {\Min} variance and the {\Pin} variance this time estimated from the full 10,000 mocks set.
We notice that the difference between the two estimates is of the order of 5\% and it is well within the estimated scatter for the variance based on 298 measurements.

In Fig.~\ref{fig:correlation} we compare instead the two correlation matrices $c_{i,j}$, defined in eq.~(\ref{eq:corr}), with the {\Min} one in the upper-left triangle and the {\Pin} one from the 10,000 mocks in the lower-right triangle. Small differences in the most off-diagonal elements are barely visible. 
 
A more quantitative comparison is shown in Fig.~\ref{fig:corr_rows} in terms of  cuts across the correlation matrix: the black dots show a row of the {\Min} correlation matrix, with the error bars estimated from the scatter of subsets of 298 realisations extracted from the 10,000 {\Pin} mocks, while the yellow curve connects the elements from the same row in the {\Pin} matrix. The 3PCF covariance matrix estimated from the 10000 {\Pin} mocks and the one from the 298 {\Min} simulations agree with each other within the uncertainty estimated for the latter. 
We also quantify the distribution of residuals of the elements of the two covariance matrices, in units of their standard deviation $\equiv \left( C_{i,j}^M-C_{i,j}^P\right)/\sigma_{C_{i,j}}$. The residuals are normally distributed, with mean $\sim0.2$ and standard deviation $\sim0.95$, close to a standard normal distribution. This deviation is indeed small with respect to the variance
and can be safely ignored.
We shall consider the {\Pin} covariance as a reference and will use it to perform all likelihood analyses presented in what follows.
%==================================================
\subsection{3PCF covariance:  {\Min} vs theory }
%==================================================
\label{sec:Minervavstheory}
 \begin{figure}
    \centering
    \includegraphics[width=0.9\textwidth]{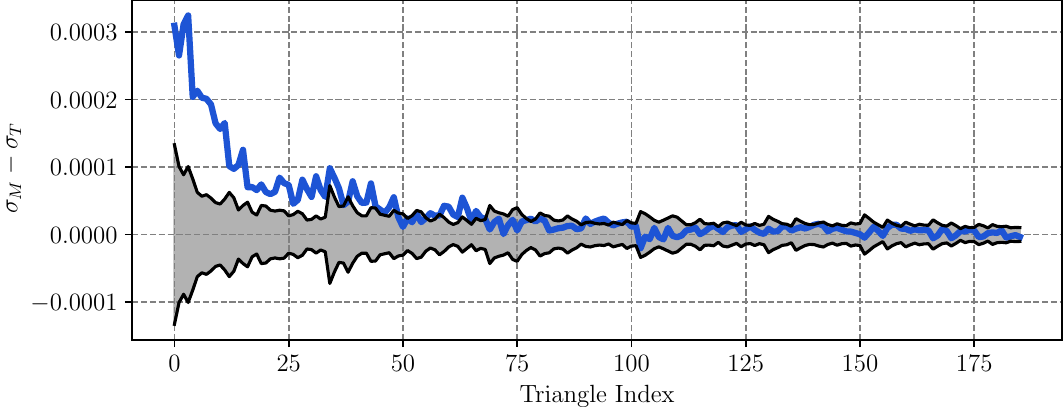}
    \label{fig:std_theo}
\end{figure}
%==================================================
We now consider an analytical model for the covariance matrix, in the Gaussian approximation, as described in \cite{Slepian2015}. An analytic expression has the obvious advantage of being noiseless and flexible to changes in the fiducial cosmology. Most importantly it does not require large sets of mock catalogs and the numerically expensive measurements of the 3PCF. On the other hand, the simple analytic model we consider here makes assumption, as the Gaussianity of the halo distribution, that could lead to potential systematic errors and does not take into account the specific geometry of the sample (but see \cite{Philcox2019} for possible improvements).

To compute the analytic covariance matrix we need to assume a fiducial, non-linear power spectrum, the covered volume and a shot noise contribution to the power spectrum. The former is provided by the best-fit model to the  {\Min} power spectrum measurements\footnote{The best-fit power spectrum has been derived and kindly provided by Kevin Pardede using the PT model and assumptions of \cite{Oddo2021}.}. The latter is set equal to the inverse of the mean halo number density in the mocks, i.e. $1./\bar{n} = 4500 \, h^{-3} \mathrm{Mpc}^3$. 

To check if the analytical, Gaussian errors are consistent with those obtained from the mock catalogs, we show in  fig.~\ref{fig:std_theo} the difference between the two sets of errors (blue curve),
compared with the 2-$\sigma$ error measured in the {\Min} mocks (grey band) obtained as in Fig.~\ref{fig:std}. We notice that the theory errors are systematically smaller than those measured from the mocks, with the mismatch decreasing with the size of the triangles. This is expected since the adequacy of the Gaussian hypothesis increases with the scale, while it is expected to break down on small scales due to the non-linear evolution inducing non-Gaussian contributions to the covariance. At larger scales the deviation is smaller, of the order of few percent. In order to reabsorb this small difference, one can exploit a  small set of measured 3PCFs to tune the theoretical covariance parameters \citep{Slepian2015, Fumagalli2022}.

%==================================================
\section{The halo 3PCF Model}
\label{sec:model}

Since one of the goals of this work is to compare the results of the 3PCF analysis to the one of the bispectrum analyses, we adopt the model of \cite{Oddo2020} and obtain the 3PCF model via inverse-FFT as for example in \cite{Jing1997, Barriga2002, Slepian2017b}. The model corresponds to the simple tree-level expression in Perturbation Theory including the contribution from gravitational instability plus the one from non-linear, local and tidal bias \cite{FryGaztanaga1993, ChanScoccimarroSheth2012, BaldaufEtal2012}. Here we follow \cite{Slepian2017b} to obtain the \textit{pre-cyclic} 3PCF part of the prediction as
\begin{align}
\label{eq:zeta_precyc}
    \zeta_{pc}(r_{12}, r_{13}, \mu_{23}) =  & b_1^3\left( \frac{34}{21}  + \frac{b_2}{b_1} - \frac{4}{3} \frac{\gamma_2}{b_1}\right) \xi_0(r_{12}) \xi_0(r_{13})+\\ \nonumber
                 & - b_1^3 \left[\xi_1^+(r_{12}) \xi_1^-(r_{13}) + \xi_1^-(r_{12}) \xi_1^+(r_{13}) \right] \mathcal{L}_1(\mu_{23}) \\
                 & + b_1^3 \left( \frac{8}{21} + \frac{4}{3} \frac{\gamma_2}{b_1} \right) \xi_2(r_{12}) \xi_2(r_{13}) \mathcal{L}_2(\mu_{23}) \nonumber \, ,
\end{align}
where the triangle sides $r_{12}, \, r_{13}, \, r_{23}$ and the cosine angle $\mu_{23}$ are related through  $r_{23}^2 = r_{12}^2 + r_{13}^2 - 2 \mu_{23} r_{12} r_{13}$ and $j_{\ell}$ are the spherical Bessel functions.

The two-point correlation function moments in eq.~(\ref{eq:zeta_precyc})
\begin{align}
    \label{eq:xil}
    \xi_{\ell} (r) & = \frac{1}{2\pi^2} \int_0^{\infty} \de k k^2 P(k) j_{\ell} \left(k r\right), \nonumber \\
    \xi_\ell^{[\pm]} (r) & = \frac{1}{2\pi^2} \int_0^{\infty} \de k k^2 k^{\pm1} P(k) j_{\ell} \left(k r\right)
\end{align}
depend on the linear matter power spectrum $P(k)$, 
that we model using \textsc{CAMB} \cite{Lewis2000}.

The full galaxy 3PCF model is then obtained by summing permutations of the $\zeta_{pc}$ function:
\begin{equation}
  \label{eq:zetah}
  \zeta_h (r_{12}, r_{13}, r_{23}) = \zeta_{pc}(r_{12}, r_{13}, \mu_{23}) + \zeta_{pc}(r_{13}, r_{23}, \mu_{12}) +
  \zeta_{pc}(r_{23}, r_{12}, \mu_{13}).
\end{equation}
To speed up the sampling of the parameter space in the likelihood analysis of Sec.~\ref{sec:paraminf} we first numerically evaluate and tabulate the integrals in eq.~(\ref{eq:xil}) in the reference $\Lambda$CDM cosmology and then
maximize the likelihood with respect to the bias parameters.

This model, however, cannot be compared directly to the quantities measured by our estimator based on the Spherical Harmonics Decomposition. We should therefore, in the first place, expand it in multipoles. In addition, we need to account for the finite size of separation bins and we do that at the multipoles level, computing
\begin{equation}
    \label{eq:zeta_ell_ave}
    \widetilde{\zeta}_{\ell}(r_{12}, r_{13}) = \frac{2l+1}{2} \int_{-1}^{1} \de \mu \mathcal{L}_{\ell}(\mu)
    \int_{r_{12}-\Delta r/2}^{r_{12}+\Delta r/2} \!\!\!\!\!\!\!\de q \, q^2
    \int_{r_{13}-\Delta r/2}^{r_{13}+\Delta r/2} \!\!\!\!\!\!\de p \, p^2\,
    \zeta(p, q, \mu) \, ,
\end{equation}
where $r_{12}, r_{13}$ represent the center of the bins, $\Delta r$ is the bin size and $\zeta(p, q, \mu) $ is the model, eq.~(\ref{eq:zetah}). In the end, the binned 3PCF model that we compare with the data is
\begin{equation}
    \label{eq:model_resum}
    \widetilde{\zeta}(r_{12}, r_{13}, r_{23}) = \sum_{l=0}^{{\ell}_{max}} \widetilde{\zeta}_{\ell}(r_{12}, r_{13})\, \widetilde{\mathcal{L}}_{\ell}(r_{12}, r_{13}, r_{23}).
\end{equation}
We notice that the comparison of the model to an accurate  estimation of the full 3PCF takes advantage of a clear dependence on the physical scales represented by the three sides. This allows in turn to a proper definition of the minimal separation $r_{min}$ at which the model fails. The accuracy of the estimation of the full 3PCF $\zeta(r_{12}, r_{13}, r_{23})$ from the equation above, however, is limited by the finite value of $\ell_{max}=10$, required if we want to preserve the efficiency of the implementation of \cite{Slepian2015}. In section~\ref{sec:results} we will explore in detail the subset of triangular configurations affected by this approximation and therefore not properly described by the tree-level expression.

%==================================================
\section{Parameter inference}
\label{sec:paraminf}

We validate the galaxy 3PCF model, in terms of the posterior distribution on the three bias parameters, testing various assumptions on the likelihood analysis.

For each catalog, identified by the $\alpha$ index, we define the $\chi^2$ function
\begin{equation}
    \label{eq:chi2}
    \chi^{2}_\alpha(\Vec{\mu}, \mathbb{C})=(\mathbf{d}_\alpha-\boldsymbol{\mu})^{T} \mathbb{C}^{-1}(\mathbf{d}_\alpha-\boldsymbol{\mu}),
\end{equation}
where $\Vec{\mu}$ and $\Vec{d}_\alpha$ represent,  respectively, the model and data vector while $\mathbb{C}$ is the covariance matrix with elements $C_{i,j}$. In the case of the (noiseless) analytic covariance matrix we adopt, for the single realization, the Gaussian likelihood function
\begin{equation}
    \mathcal{L}_\alpha \propto \exp\left(- \frac{\chi^2_\alpha}2  \right)\, .
\end{equation}
We will also consider covariance matrices, $\widehat{\mathbb{C}}$, estimated numerically from a limited number of $N_m$ halo catalogs, whose statistical noise can bias the determination of the precision matrix $\mathbb{C}^{-1}$. We then include the correction suggested in \cite{Hartlap2007} and given by
\begin{equation}
    \mathbb{C}^{-1}=\frac{N_m-n_d-2}{N_m-1}\,\widehat{\mathbb{C}}^{-1}\,
\end{equation}
where $N_m$ is the number of independent measurements used to estimate $\widehat{\mathbb{C}}$ while $n_d$ is the size of the data vector. The correction inflates the errors in order to encompass any possible systematics on $\mathbb{C}^{-1}$. In the case of the covariance estimated from the full set of 10,000 {\Pin} mocks, given a data vector comprising at most 184 triangular configurations, the correction is rather small and negligible. This is not the case, naturally, for the covariance estimated from the more limited set of {\Min} simulations, where the factor can be as low as 0.35. We considered as well the prescription of  \cite{Sellentin2016} finding marginal differences in our applications.

The total likelihood adopted for each evaluation is obtained as the product of all individual likelihood, or
\begin{equation}
    \label{eq:likTot}
    \log{\mathcal{L}_{tot}} = \sum_{\alpha=1}^{N_{m}} \log{\mathcal{L}_\alpha}\,.
\end{equation}

Finally, we use the correction factor in \cite{Percival2014} to fully take into account the propagation of covariance matrix uncertainty on the parameter posterior distributions. The factor is
\begin{equation}
    \label{eq:cov_percival}
    m_1 = \frac{1+B(n_d-n_p)}{1+A+B(n_p+1)}; 
\end{equation}
with $n_p$ the size of the model parameter vector, and 
\begin{align}
    \label{eq:percival_AB}
    A & = \frac{2}{(N_m - n_d -1)(N_m-n_d-4)}, \\
    B & = \frac{N_m - n_d -2}{(N_m - n_d -1)(N_m-n_d-4)}\,. 
\end{align}

We adopt flat, uninformative priors,  specified in Table~\ref{tab:summary_fit}. To sample the likelihood function we use a Monte Carlo Markov Chain (MCMC) approach as implemented in the \textsc{emcee} software \citep{ForemanMackey2013}.

%==================================================
\section{Results}
\label{sec:results}

\begin{table}[]
    \centering
    \renewcommand{\arraystretch}{1.4}
    \begin{tabular}{c|c|c|c|c|c|c}
         \hline
         Name & Binning & Covariance & $b_1$ & $b_2$ & $\gamma_2$ & Sec.\\
         \hline
         \textit{Reference} & Integrated & {\Pin} & $\mathcal{U}(0.5, 5)$ & $\mathcal{U}(-10, 10)$ &
         $\mathcal{U}(-10, 10)$ &
         \ref{sec:scale_sens}\\
         \textit{Benchmark} & Integrated & \textsc{\Min} & $\mathcal{U}(0.5, 5)$ & $\mathcal{U}(-10, 10)$ &
         $\mathcal{U}(-10, 10)$ &
         \ref{sec:minerva_cov}\\
         \textit{Gaussian} & Integrated & Theoretical & $\mathcal{U}(0.5, 5)$ & $\mathcal{U}(-10, 10)$ &
         $\mathcal{U}(-10, 10)$ &
         \ref{sec:theo_cov}\\
         
         \textit{Shrinkage} & Integrated & Shrinkage & $\mathcal{U}(0.5, 5)$ & $\mathcal{U}(-10, 10)$ & $\mathcal{U}(-10, 10)$ & \ref{sec:shrinkage}\\
         \textit{Bin-center} & Center & {\Pin} & $\mathcal{U}(0.5, 5)$ & $\mathcal{U}(-10, 10)$ &
         $\mathcal{U}(-10, 10)$ &
         \ref{sec:binning}\\  
         $\gamma_2(b_1)$ & Integrated & {\Pin} & $\mathcal{U}(0.5, 5)$ & $\mathcal{U}(-10, 10)$ &
         Eq.~\ref{eq:gamma2_b1} &
         \ref{sec:bias}\\
         $b_2(b_1)$ & Integrated & {\Pin} & $\mathcal{U}(0.5, 5)$ &  Eq.~\ref{eq:b2_b1}&
         $\mathcal{U}(-10, 10)$ &
         \ref{sec:bias}\\  
         \hline
    \end{tabular}
    \caption{Summary of test case and their main characteristics. }
    \label{tab:summary_fit}
\end{table}
%==================================================
In this section we will consider the effect of different triangle selections, covariance matrices and binning schemes, as well as the inclusion of relations among bias parameters to reduce the parameter space. Table~\ref{tab:summary_fit} lists the characteristics of the various tests and summarizes the methodological choices adopted in each of them. All runs share the same {\Min} halo catalogs at $z=1$ where the 3PCF is measured with the SHD estimator using $\ell_{max}=10$, $\Delta r=10 \hMpc$ and $\eta_{min} = 1$.
The Gaussian likelihood is maximized with respect to the three halo bias parameters $b_1, \, b_2$ and $\gamma_2$. All other parameters are kept fixed.

%==================================================
\subsection{Sensitivity to triangle size and shape}
\label{sec:scale_sens}

\begin{table}[]
\centering
\renewcommand{\arraystretch}{1.4}
\begin{tabular}{cllllllll}
\multicolumn{1}{l}{}                               &                        & \multicolumn{7}{c}{$r_{min} \, [\hMpc]$}                                                                                                                                          \\ \cline{3-9} 
\multicolumn{1}{l}{}                               & \multicolumn{1}{l|}{}  & \multicolumn{1}{l|}{10} & \multicolumn{1}{l|}{20} & \multicolumn{1}{l|}{30} & \multicolumn{1}{l|}{40} & \multicolumn{1}{l|}{50} & 60                    & \multicolumn{1}{l|}{70} \\ \cline{2-9} 
\multicolumn{1}{c|}{\multirow{3}{*}{$\eta_{min}$}} & \multicolumn{1}{l|}{1} & \multicolumn{1}{l|}{216}   & \multicolumn{1}{l|}{186}   & \multicolumn{1}{l|}{152}   & \multicolumn{1}{l|}{117}   & \multicolumn{1}{l|}{84}    & \multicolumn{1}{l|}{56}   & \multicolumn{1}{l|}{35}   \\
\multicolumn{1}{c|}{}                              & \multicolumn{1}{l|}{2} & \multicolumn{1}{l|}{170}   & \multicolumn{1}{l|}{143}   & \multicolumn{1}{l|}{113}   & \multicolumn{1}{l|}{83}    & \multicolumn{1}{l|}{56}    & \multicolumn{1}{l|}{35}   &  \multicolumn{1}{l|}{20}   \\
\multicolumn{1}{c|}{}                              & \multicolumn{1}{l|}{3} & \multicolumn{1}{l|}{131}   & \multicolumn{1}{l|}{107}   & \multicolumn{1}{l|}{81}    & \multicolumn{1}{l|}{56}    & \multicolumn{1}{l|}{35}    & \multicolumn{1}{l|}{20}    & \multicolumn{1}{l|}{10}   \\ \cline{2-9} 
\end{tabular}
\caption{Number of triangles as a function of $\eta_{min}$ and $r_{min}$. $r_{max}$ is fixed to $130 \, \hMpc$.}
\label{tab:nt_table}
\end{table}

\begin{figure}
\centering
\includegraphics[width=\textwidth]{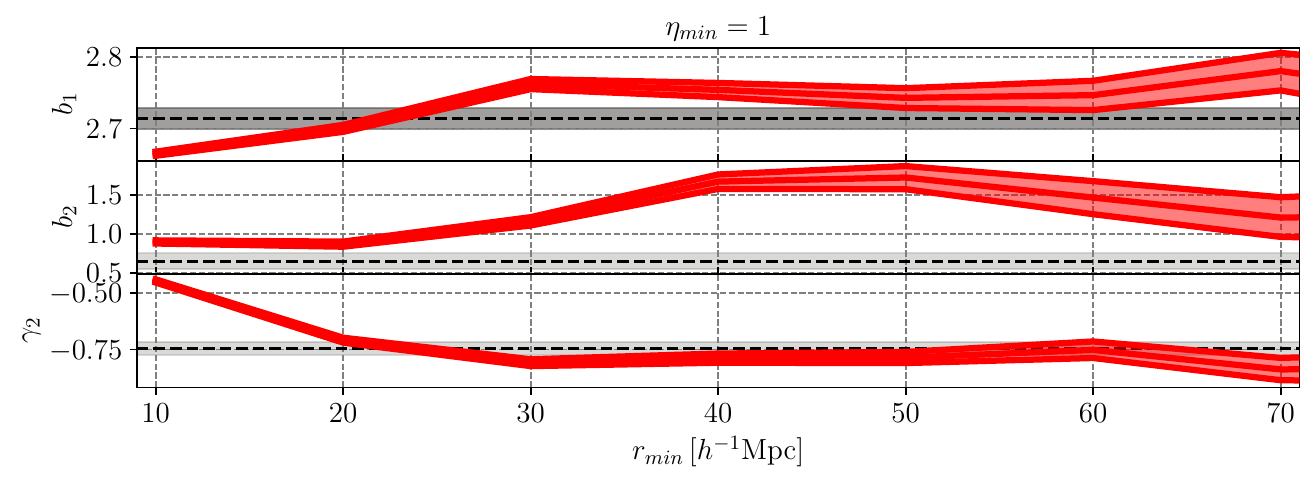} \\
\includegraphics[width=\textwidth]{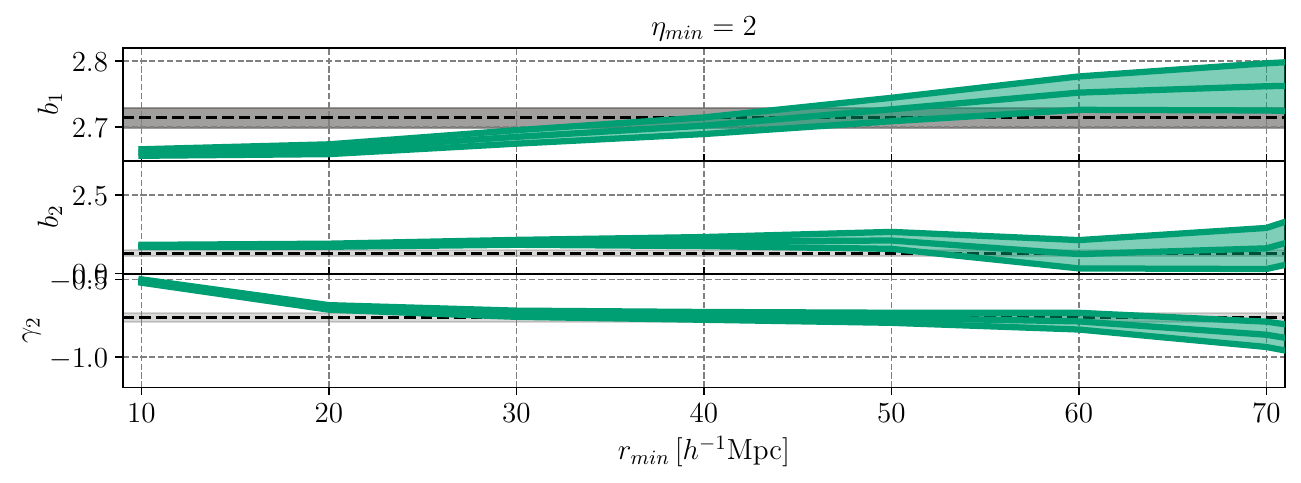} \\
\includegraphics[width=\textwidth]{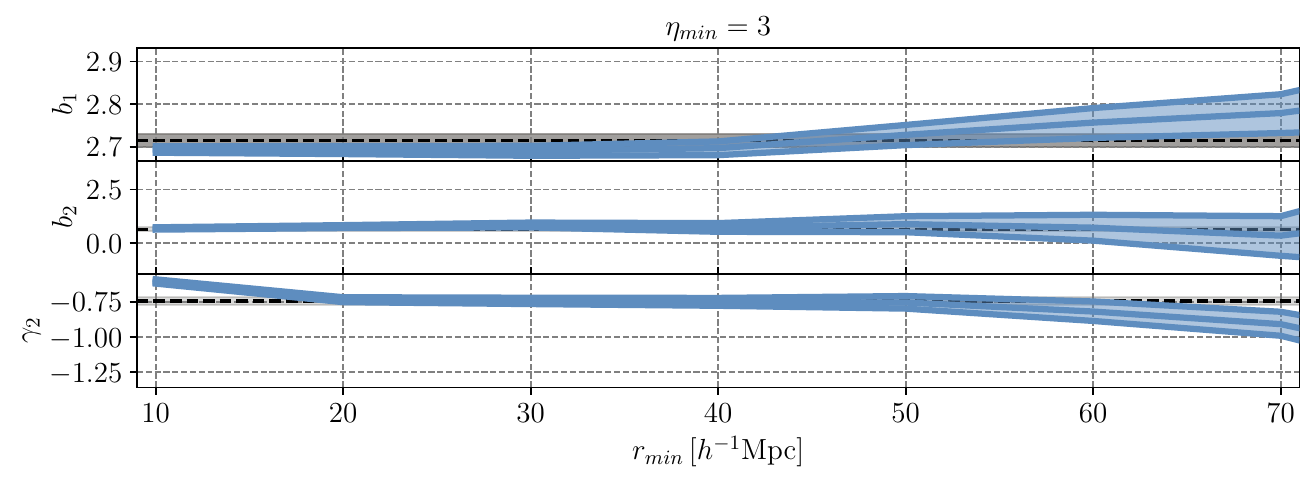}
\caption{Inferred bias parameters 
as a function of the minimum triangle side
$r_{min}$ (thick red curve) with their  $1\sigma$ uncertainty (shaded area) for the Reference case. Dashed black lines with
grey shaded region represent the bias values and their $1\sigma$ errors estimated from the bispectrum analysis of \cite{Oddo2020}. \textit{Top, central and bottom panels}: all results are for the reference case using triangles with $\eta\geq1, \, 2$ and 3, respectively.} 
\label{fig:benchmark_scale}
\end{figure}
%==================================================

\begin{figure}
\centering
\includegraphics[width=0.8\textwidth]{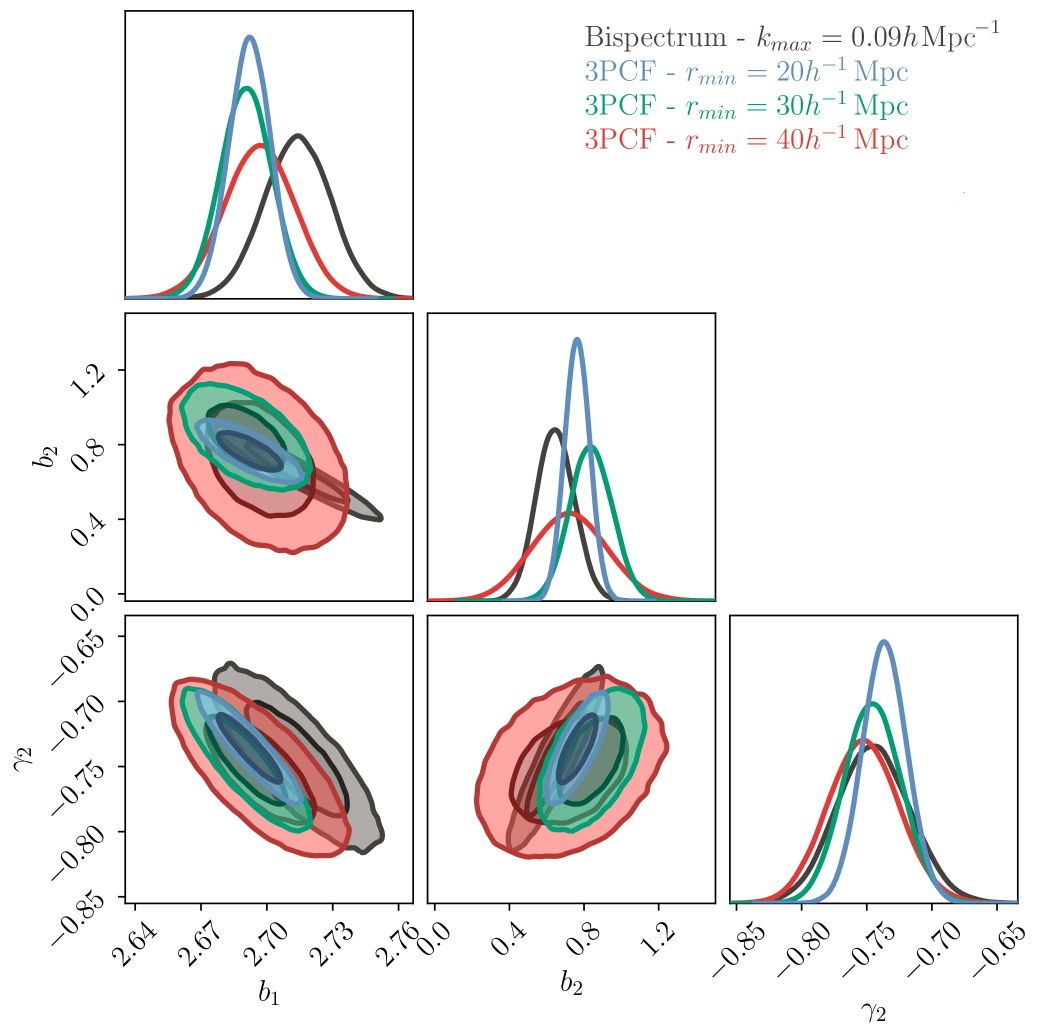}
\caption{$1D - 2D$ posterior probability distributions of the bias parameters 
inferred from the likelihood analysis in the reference case. The green, blue and red contours (curves) are used for the cases  the cases $r_{min} = 20, 30, 40 \, \hMpc$ and $\eta_{min}= 3$. Reference dark grey contours (curves) are from the bispectrum analysis of \cite{Oddo2020}.
}
\label{fig:benchmark_cntrs}
\end{figure}

To assess the sensitivity of the 3PCF analysis to the size and shapes of the triangles we performed different likelihood analyses in which we fix $\eta_{min}$, hence selecting the triangle types, and progressively increase the value of $r_{min}$. The number of triangles that remain after a specific selection is shown in Tab.~\ref{tab:nt_table}.

The results are shown in Fig.~\ref{fig:benchmark_scale}. In each panel we plot the best fit bias parameters ($b_1$, $b_2$, $\gamma_2$, from top to bottom) as a function of $r_{min}$. Shaded areas represent 1-$\sigma$ uncertainty. We also plot, for reference, the best fit values (and uncertainties) obtained by \cite{Oddo2020} from the bispectrum analysis of the same dataset with Fourier space triangles smaller than $k_{max}=0.082 \kMpc$ (dashed black lines and grey shaded areas). Each panel shows the results obtained with $\eta_{min} = 1, 2$ and $3$  (red, green and blue curves, respectively), corresponding to excluding those triangles with no or a small difference between the sides $r_{12}$ and $r_{13}$, see eq.~(\ref{eq:eta}).

Let us focus first on the cases with $\eta_{min}= 2 $ and 3. In these cases, both parameters $b_2$ and $\gamma_2$, for $ r_{min} \geq 20\, \hMpc$, converge to a constant value consistent with the one obtained in the bispectrum analysis. Only the linear bias parameters $b_1$ exhibits a scale dependence, particularly significant for $ r_{min} $ up to $\sim 40 \hMpc$ and  for $\eta_{min}=2$.

The case of $\eta_{min}= 1$ is different. In this case $\gamma_2$ and $b_1$ show some dependence on $r_{min}$; the former converges to the correct value, but we should notice how the behavior of $b_2$ shows no clear trend and it is inconsistent with the bispectrum results for all values of $r_{min}$. This can be explained recalling that the case $\eta_{min}=1$ includes the \textit{quasi-isosceles} configurations, that is those with $r_{12}\simeq r_{13}$, with the third side $r_{23}$ approaching zero. When this happens, the 3PCF tree-level model
that we adopted in our analysis, specific to the fast estimator described in section~\ref{sec:zetameas}, fails \cite{Slepian2015, Veropalumbo2021}. In fact, both the estimator and the model are defined by a maximum value of $\ell_{max}=10$ in the Legendre expansion, eq.~(\ref{eq:zeta_ell}), and probably a larger value of $\ell_{max}$ would have reduced the dependence on $\eta_{min}$. In the rest of this work we adopt $\eta_{min}= 3$ for all our results. 

To inspect the degeneracy between the bias parameters and how this depends on $r_{min}$, we plot their 1D and 2D marginalized distributions in the triangle plot of Fig.~\ref{fig:benchmark_cntrs}.
We compare the three cases of $r_{min}=20, 30$ and $40 \hMpc$ (respectively in blue green and red) with the result of the bispectrum analysis (grey) from \cite{Oddo2020}, where a scale cut of $k_{max}=0.09\kMpc$ is assumed. 

For $r_{min} \leq 30 \hMpc$ the constraints from the 3PCF analysis are slightly tighter and less degenerate than in the bispectrum case, with the $b_1$ value recovered from the 3PCF 1-$\sigma$ lower than the one from the bispectrum. On the other hand, the error on $b_2$ inferred from the bispectrum is smaller. 
The posteriors width is a non-trivial combination of two aspects: the ability of the different probes to break degeneracies among parameters and the information content as a function of the scale. It is indeed evident, in Fig.~\ref{fig:benchmark_cntrs}, that the two probes are not equally able to constrain bias parameters, particularly $b_2$, with the difference showing a notable dependence on the scales-cut.
Yet, the overall agreement is an indication that our understanding of halo 3-point correlation in terms of local and tidal bias along with the control of systematic in both the Fourier-space as the configuration-space estimations and analyses are on rather solid grounds. A more detailed comparisons deserve a dedicated discussion that is beyond the aims of this paper.

%==================================================

\subsection{Noisy estimates of the numerical covariance}
\label{sec:minerva_cov}

In Section~\ref{sec:MinervavsPinocchio} we found no significant difference between the 3PCF covariance matrices estimated from the {\Min} simulations and the {\Pin} approximate mocks. We can nevertheless expect a bias, particularly  on the inverse covariance matrix, when the numerical estimation is based on a small number of catalogs. As described in Sec.~\ref{sec:paraminf}, we account for this enlarging the error bars to encompass possible systematic effects.

\begin{figure}[t!]
    \centering
    \includegraphics[width=0.9\textwidth]{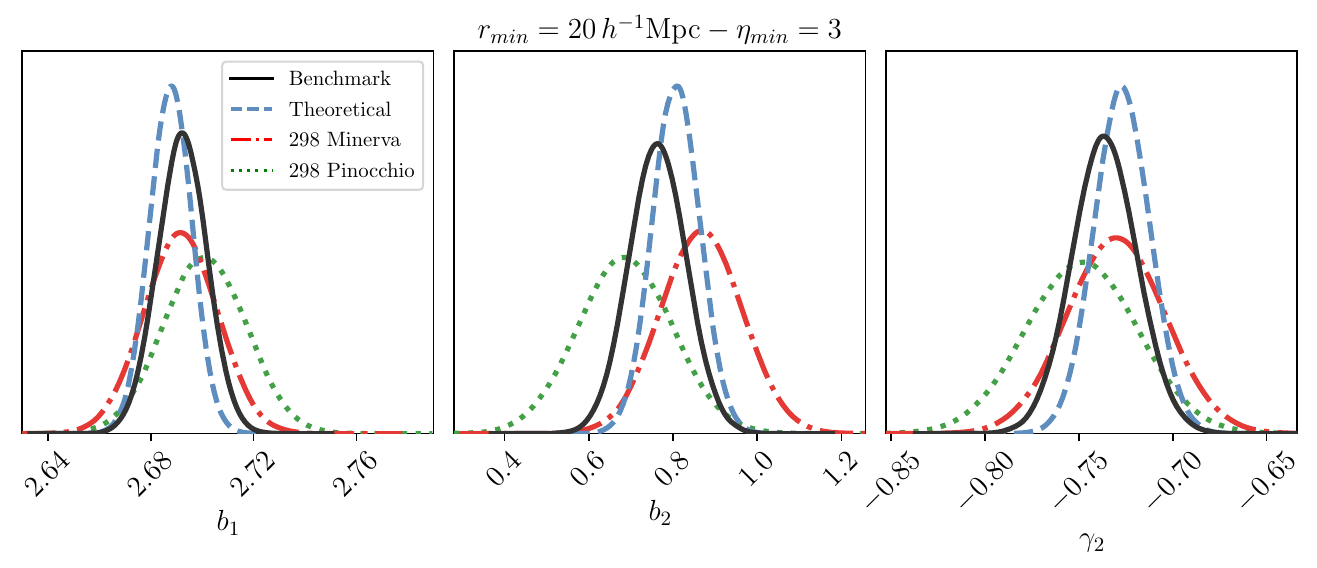} \\
     \includegraphics[width=0.9\textwidth]{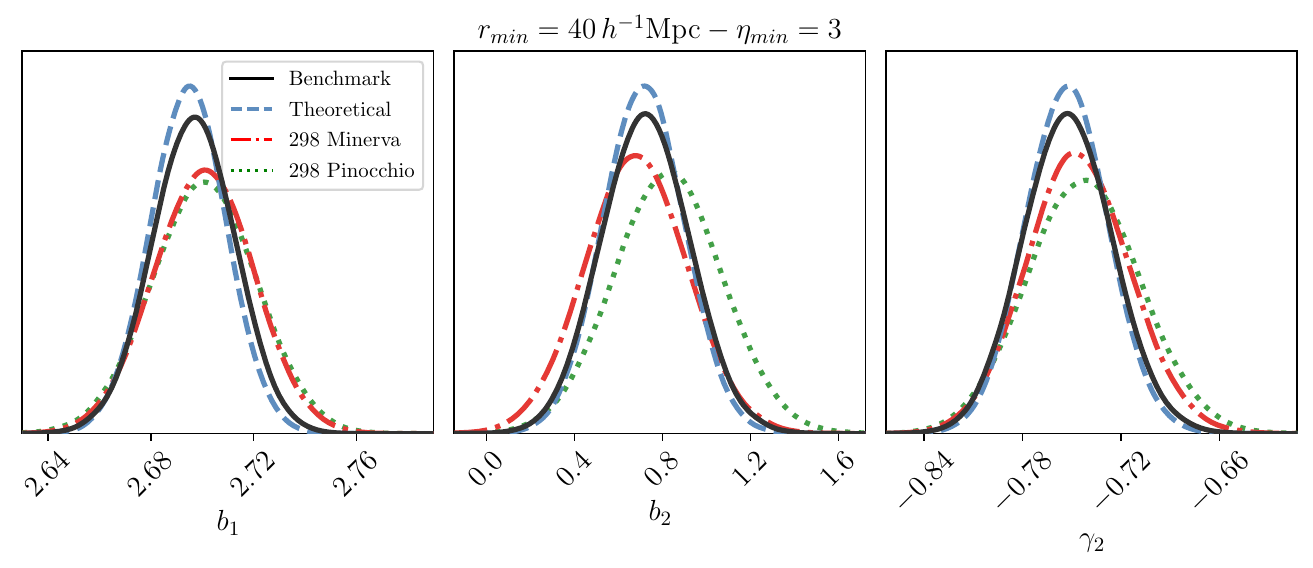} \\
    \caption{Marginalized posterior distribution of three bias parameters assuming $\eta_{min}=3$, $r_{min}=20$ (top panels) and $r_{min}=40$ (bottom panels), obtained from different covariance matrices: numerical estimates from the full {\Pin} set (black continuous curve), small set of 298 catalogs from the {\Min} simulations (red dot-dashed), 298 catalogs from the {\Pin} mocks (green dotted) and the analytic Gaussian model (blue dashed).}
    \label{fig:bias_cov}
\end{figure}

In Fig.~\ref{fig:bias_cov} we compare the marginalized posteriors on the bias parameters obtained assuming the covariance matrix estimated from the small sets of 298 {\Min} simulations (red curves) and the one estimated from the same number of {\Pin} mocks (green) to the reference case of the covariance from the full {\Pin} set. (black). The top panels assume $r_{min}=20\hMpc$, while for the lower ones the data vector is limited to $r_{min}=40\hMpc$. For $r_{min}=20\hMpc$ the main difference is clearly due to the correction factor applied to the poorly estimated covariances. In this case we notice, in addition, a  shift of about 1-$\sigma$ in the mean of the $b_2$, sign of a systematic error induced by a poor estimation of the covariance. 
We also notice a shift in the mean values between the 298 {\Min} and 298 {\Pin} cases, despite they share the same initial conditions. This suggests that noisy covariance obtained with approximated methods should be taken with care.
This discrepancy disappears when excluding triangles with sides smaller that 40 $\hMpc$ from the analysis. Similar results are obtained with different values  for $\eta_{min}$ values and for larger $r_{min}$ sizes and for this reason they are not shown here.

%==================================================
\subsection{Analytical covariance model}
\label{sec:theo_cov}

\begin{figure}
    \centering
    \includegraphics[width=0.85\textwidth]{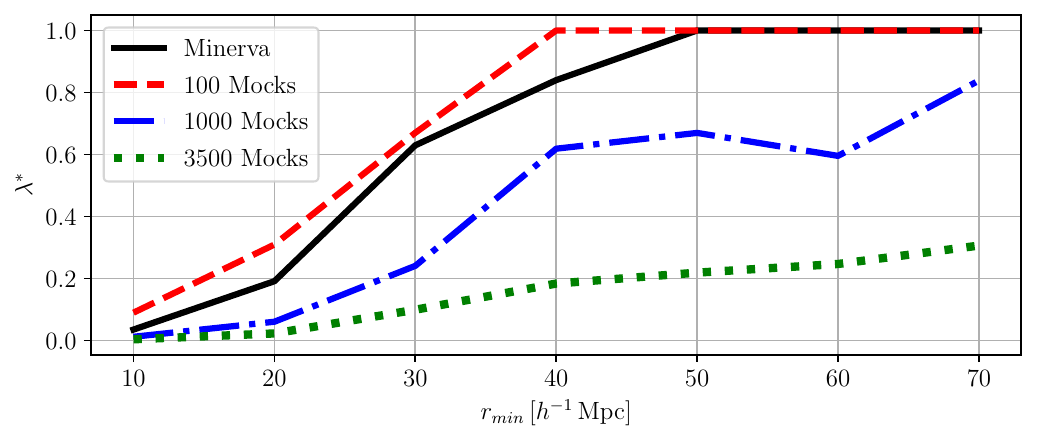}
    \caption{Value of the shrinkage intensity $\lambda^*$ as a function of the minimum triangle scale. Different curves correspond to different choices of the numerical matrix $C_{i,j}$ estimated from 298 {\Min} simulations (black, continuous) and 100, 1000 and 3500 {\Pin} mocks (respectively red dashed, blue dot-dashed and green dotted). For this figure we select only triangles with $\eta_{min}=3$.} 
    \label{fig:lambdas}
\end{figure}

We turn now to the effect of the analytical covariance on the on the likelihood analysis. The resulting posterior distributions of the bias parameters are shown as blue dashed curves in Fig.~\ref{fig:bias_cov}. Not surprisingly, in the  $r_{min}=20\,\hMpc$ case (top panels) the marginalized uncertainties are underestimated w.r.t. the case of the reference numerical covariance. However, the difference is not very large and we do not notice a significant bias on the central values of the parameters. The comparison improves significantly for $r_{min}=40\,\hMpc$ (bottom panels), indicating that the Gaussian hypothesis is rather adequate for 3PCF analyses above this scale.

%==================================================
\subsection{Shrinkage covariance matrix}
\label{sec:shrinkage}

Several methods have been proposed to estimate the covariance matrix with a limited number of mocks while keeping a good accuracy \cite{Pope2008, PazSanchez2015}. We consider here the shrinkage technique of \citep{Pope2008}, based on the combination of a noisy but unbiased estimate of the covariance (e.g. from N-body simulations) with a possibly biased but more precise estimate or noiseless model (e.g. from theory). The benefit of this combination is that of reducing the number of mocks but also, in our case, the non-negligible computational cost of estimating their 3PCFs.

In practice, we consider in the linear combination the (biased but noiseless) theoretical, Gaussian covariance $T_{i,j}$ with the numerical covariance, $\widehat{C}_{i,j}$, that is
\begin{equation}
    \label{eq:shrinkagecov}
    S_{i,j} = \lambda^*T_{i,j} + (1-\lambda^*)\widehat{C}_{i, j}\,,
\end{equation}
where the parameter $\lambda^*$ represent the \textit{shrinkage intensity}. For $\lambda^*=0$ one recover the numerical covariance while for $\lambda^*=1$ we have the theory model. The optimal choice for $\lambda$ is found by minimizing a mean-square difference cost function giving, in our case
\begin{equation}
    \label{eq:lambdastar}
    \lambda^{\star}=\frac{\sum_{i, j} \widehat{\operatorname{Var}}\left(\widehat{C}_{i j}\right)}{\sum_{i, j}\left(T_{i j}-\widehat{C}_{i j}\right)^{2}}\,,
\end{equation}
where $\widehat{\operatorname{Var}}\left(\widehat{C}_{i j}\right)$ is the variance associated with each $\widehat{C}_{i,j}$, that we compute as \citep{Pope2008}
\begin{equation}
    \label{eq:VarCov}
    \widehat{\operatorname{Var}}\left(\widehat{C}_{i j}\right)=\frac{N_m}{(N_m-1)^{3}} \sum_{\alpha=1}^{N_m}\left(W_{i j}^{(\alpha)}-\widehat{W}_{i j}\right)^2\,,
\end{equation}
with
\begin{equation}
    \label{eq:W}
W_{i j}^{(\alpha)}=\left(\hat{\zeta}_{i}^{(\alpha)}-\overline{\zeta}_{i}\right)\left(\hat{\zeta}_{j}^{(\alpha)}-\overline{\zeta}_{j}\right)\,
\end{equation}
computed in terms of the 3PCF measurement $\hat{\zeta}_{i}^{(\alpha)}$ for the $i$-th triangle in the $\alpha$-th realization and the average $\overline{\zeta}_{i}$ at triangle $i$, and with
\begin{equation}
    \label{eq:Wave}
\widehat{W}_{i j}=\frac{1}{N_m} \sum_{\alpha=1}^{N_m} W_{i j}^{(\alpha)}\,.
\end{equation}

We show our estimates of  $\lambda^*$ as a function of $r_{min}$, assuming $\eta_{min}=3$, in fig.~\ref{fig:lambdas}. Different curves correspond to different choices of the numerical matrix $\widehat{C}_{i,j}$. As expected, on large scales $\lambda^* \rightarrow 1$ since the covariance is approximately Gaussian, although when 1000 or more mocks are available, the parameter $\lambda^*$ never reaches unity, showing relevant differences between the numerical estimate and analytical predictions at all scale. At small scales, where deviations from Gaussianity are larger, the numerical covariance is always favoured. In the {\Min} case, we also run the likelihood analysis with the shrinkage covariance. Unsurprisingly, the same trend found for $\lambda^*$ reflects on the constraints on the bias parameters, which stay close to the constraints shown in Fig.~\ref{fig:bias_cov}.

%==================================================

\subsection{Sensitivity to binning}
\label{sec:binning}
\begin{figure}
    \centering
    \includegraphics[width=\textwidth]{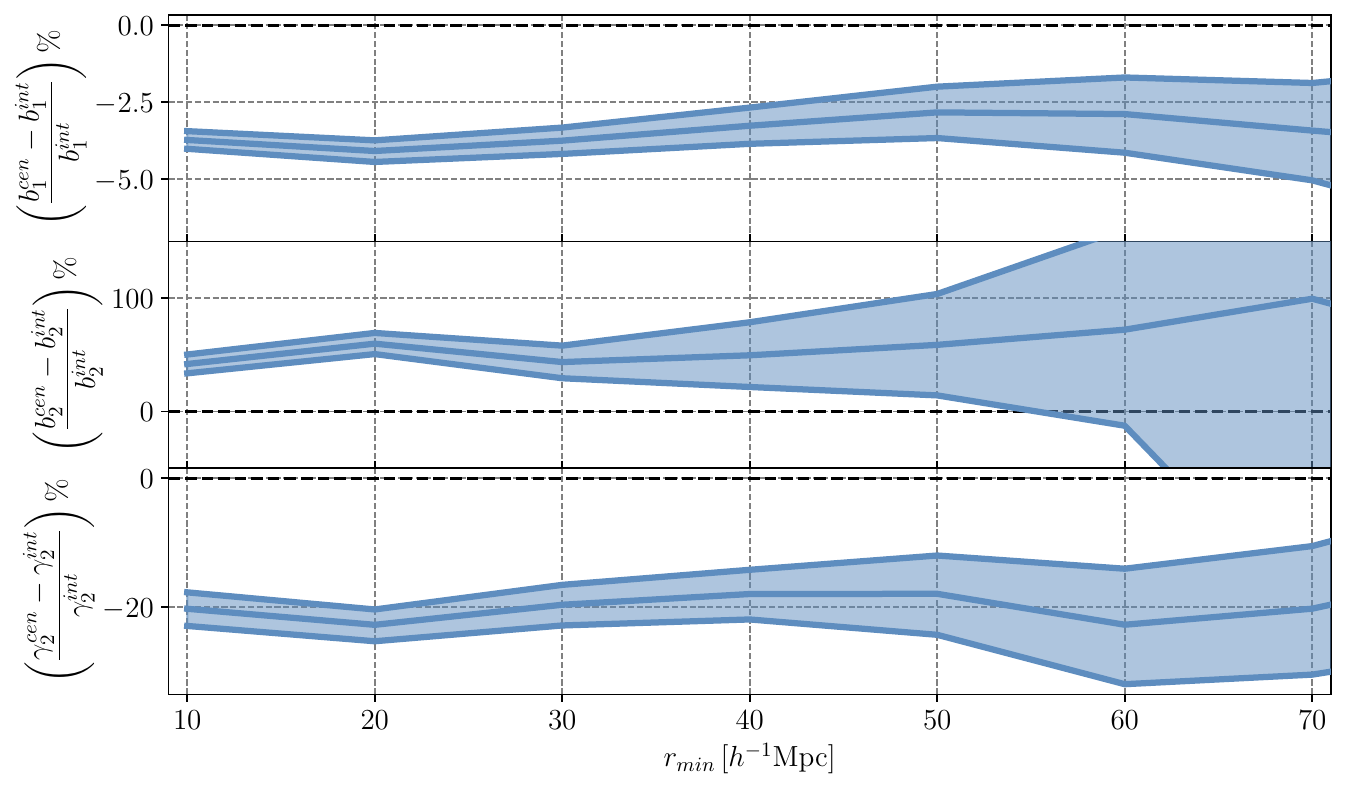}
    \caption{Percentage difference between bias constraints obtained with and without averaging over the triangle side bin, as a function of $r_{min}$ for $\eta\geq3$. The black dashed line drawn at the 0 value (perfect match)  indicates when the two hypothesis give consistent constraints.}
    \label{fig:bin_center_bias}
\end{figure}

In all the results presented so far we have used the 3PCF multipole model predictions obtained by performing the three-dimensional integral in eq.~(\ref{eq:zeta_ell_ave}). This step significantly contributes to the total computing budget and becomes particularly burdensome when performed at each likelihood evaluation in a Markov Chain exploring cosmological parameters. An alternative, cheaper option could be the simple angle integration
\begin{equation}
    \label{eq:zeta_ell_center}
    \zeta_{\ell}(r_{12}, r_{13}) = \frac{2\ell+1}{2} \int_{-1}^{1} \de \mu \mathcal{L}_{\ell}(\mu) \zeta(r_{12}, r_{13}, r_{23})  ,
\end{equation}
with $r_{12}$ and $r_{13}$ evaluated at the center of the separation bin.

In Fig.~\ref{fig:bin_center_bias} we compare the results of the likelihood analysis obtained in these two cases. We show, for each bias parameters, the percent difference relative to the reference case in which the binning effect has been fully modelled, along with the relative marginalized uncertainty. The simplification given by the evaluation at the bin center can induce significant deviations from the reference values, as large as 50\% in the case of $b_2$, and for all values of $r_{min}$ up to 70$\hMpc$.

We acknowledge the fact that the effect may be particularly large due to the bin size of $\Delta r = 10 \, \hMpc$, and that using finer binning will reduce the mismatch. That said, our results show that the analysis is sensitive to the binning adopted and that an alternative definition eq.~(\ref{eq:zeta_ell_center}) to speed up the analysis should be considered carefully.

%==================================================

\subsection{Bias relations}
\label{sec:bias}
\begin{figure}
    \centering
    \includegraphics[width=\textwidth]{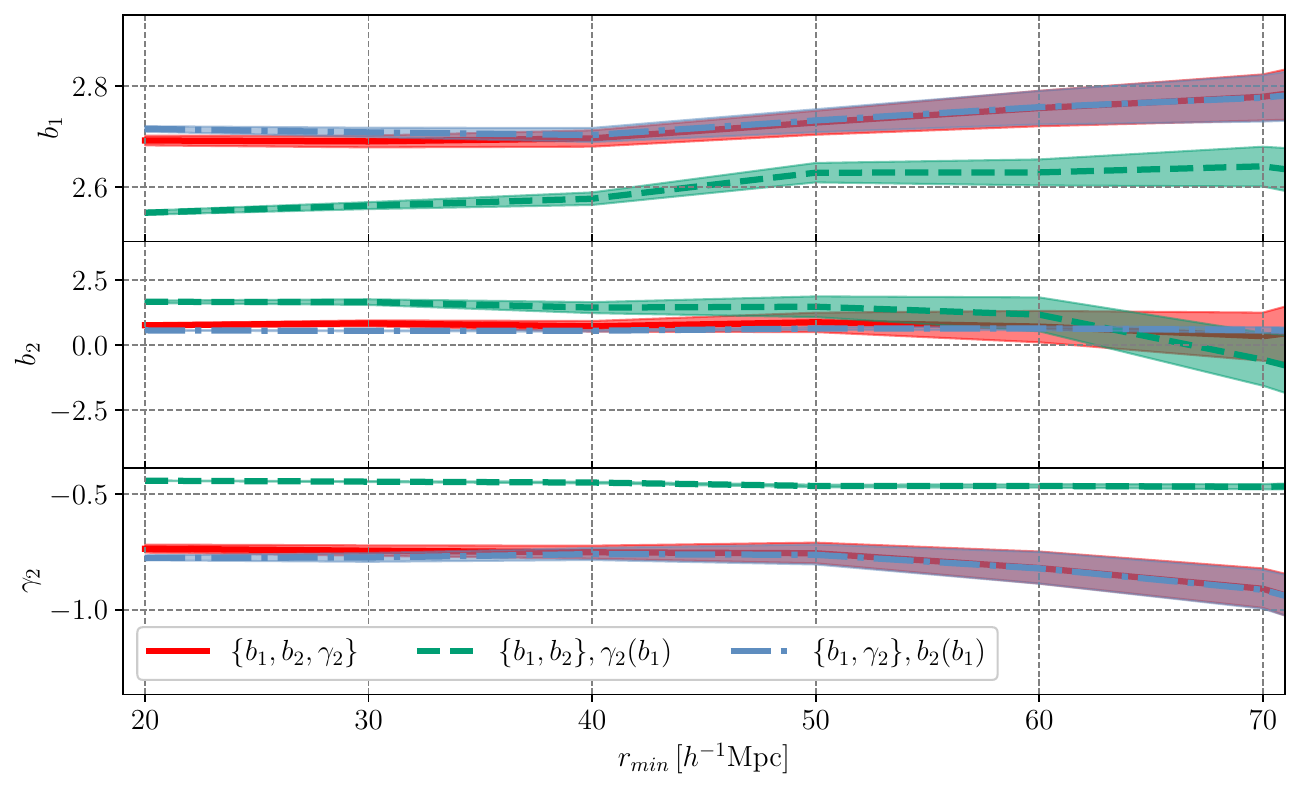}
    \caption{Best fit values of the bias parameters as a function of $r_{min}$ obtained in the reference case (red continuous curve), assuming the LLB hypothesis (dashed green curves) and using the N-body calibrated fitting function (blue, dot-dashed curve). Dashed areas represent the 1-$\sigma$ uncertainty interval. All results are obtained for triangles with $\eta\geq3$.
} 
    \label{fig:bias_rel_compare}
\end{figure}

So far we assume all three bias parameters free to vary.
However, these parameters are  not fully independent, and several relations among them can be found either from theoretical considerations or as fits to numerical simulations. Here we consider specifically the local Lagrangian bias (LLB) prescription for the tidal parameter \citep{ChanScoccimarroSheth2012, BaldaufEtal2012}
\begin{equation}
    \label{eq:gamma2_b1}
    \gamma_2(b_1) = -\frac{2}{7}(b_1-1)\, ,
\end{equation}
and the numerical fit \citep{Lazeyras2016}:
\begin{equation}
    \label{eq:b2_b1}
    b_2(b_1) = 0.412 - 2.143\, b_1 + 0.929\, b_1^2 + 0.008\, b_1^3 + \frac{4}{3}\gamma_2\, .
\end{equation}

Fig.~\ref{fig:bias_rel_compare} shows the mean over marginalized posterior of bias parameters of the best fit and their 1-$\sigma$ uncertainties as a function of $r_{min}$ for the case of $\eta_{min}= 3$. The results obtained assuming the LLB prescription (dashed green curves) are not compatible with those obtained in the reference case (red continuous curves), especially for the derived $\gamma_2$ parameter, confirming previous results \cite{Lazeyras2018, Abidi2018, Oddo2020}. On the other hand, the posteriors obtained assuming the N-body-calibrated relation in eq.~(\ref{eq:b2_b1}) (dot-dashed blue curves) are in very good agreement with the reference ones for all $r_{min}$ values. A similarly good agreement has been found in the bispectrum analysis of \cite{Oddo2020}.

%==================================================
\section{Discussion and conclusions}
\label{sec:concl}

This work extends to the 3-point correlation function the detailed analysis, previously presented for the power spectrum and bispectrum in \cite{Oddo2020, Alkhanishvili2021, ByunEtal2021, Oddo2021, PardedeEtal2022A, RizzoEtal2022A}, of a large set of 298 halo catalogs from the {\Min} simulations \cite{Grieb2016}, corresponding to a total volume of about 1,000 $\cGpc$. These are supplemented with an even larger set of 10,000 approximate halo catalogs obtained with the {\Pin} code on the same box and cosmology and reproducing the large-scale amplitude of the simulations' halo power spectrum. While not directly corresponding to any galaxy population from actual surveys, these large data-sets allow for rigorous test of several methodological choices in the analysis of galaxy clustering, taking advantage of a robust estimate of their covariance properties, even in the case of higher-order statistics.

We measure the real-space 3PCF from all the {\Min} halo catalogs as well as from the {\Pin} mocks. Each measurement includes all measurable triangular configurations with sides between 20 and 130$\hMpc$, with only a small subset excluded due to the limitations of the fast 3PCF estimator of \cite{Slepian2015}. This is crucial to explore the potential of the 3PCF in constraining bias and cosmological parameters. In our tests, we focus on the determination of the three bias parameters required by the tree-level model in Perturbation Theory. 

\begin{itemize}

    \item We explored how the recovered bias parameter depend on the value of the smallest separation $r_{min}$ included in the analysis as well as on the minimal, relative difference $\eta_{min}$ between the sides $r_{12}$ and $r_{13}$, defined in eq.~(\ref{eq:eta}). For $\eta_{min}=1$ the sum of eq.~(\ref{eq:model_resum}) limited to $\ell_{max}=10$ does not provide an accurate estimate of the full 3PCF $\zeta$ and consequently is not properly described by the tree-level model. We find that setting $\eta_{min}= 3$ leads to a consistent determination of the linear bias $b_1$ and the non-linear parameters $b_2$ and $\gamma_2$, with values independent  of $r_{min}$ for $r_{min}\ge 20 \, \hMpc$. Some dependence of $b_1$ on $r_{min}$ is evident instead for $\eta_{min}=2$, while $\eta_{min}=1$ leads to largely inconsistent results for all bias parameters. 

    \item We compare our constraints on the bias parameters for $\eta_{min}=3$ (assumed as reference value in all that follows) to the same constraints obtained from the bispectrum analysis of \cite{Oddo2020}. We find a remarkable agreement, particularly for the scale cuts given by $r_{min}=40 \hMpc$ and $k_{max}=0.09\kMpc$ for the 3PCF and bispectrum, respectively. This validates both the theoretical description of the large-scale 3-point statistics in terms of local and tidal bias contributions as well as our accurate estimation and modelling. This is, as far as we are aware, the most balanced comparison between the two statistics, based, on both sides, on all measurable configurations and on a robust numerical estimate of the covariance from a very large set of mocks.
    
    \item We compare the results obtained from covariance matrix estimated from the {\Pin} mocks to those assuming a covariance obtained from more limited set of {\Min} N-body simulations (which also provide our data-vector), accounting for possible systematic errors due to poor statistics in terms of the factor proposed by \cite{Hartlap2007}. The results are very similar for $r_{min}=40\hMpc$, while we notice some systematic shift in the determination of $b_2$ for $r_{min}=20\hMpc$, possibly due to small-scale suppression of power characteristic of the Lagrangian PT on which {\Pin}, as many other similar approximate methods, are based.
   
    \item We show a comparison of the same results with those obtained using an analytical model for the covariance matrix, assuming the Gaussian approximation  \cite{Slepian2015}. Again we find largely consistent results for $r_{min}\geq 40\, \hMpc$, while including smaller scales we can notice a slight underestimate of the marginalized uncertainty for all parameters, w.r.t. the reference numerical covariance. This is expected also from a direct comparison in terms of the variance alone, with the analytical model providing lower values at small scales.

    \item We investigate how the shrinkage technique \cite{Pope2008} can provide an hybrid estimate of the covariance based on the possibly biased analytical model and on a noisy numerical estimate, determining the shrinkage parameter $\lambda^*$ as a function of $r_{min}$ for different choices of the numerical matrix. We find that in all cases the numerical result is preferred when $r_{min} \lesssim 30\hMpc$, but also that a significant numerical component is retained when the number of available mocks is sufficiently large ($\sim 1000$). 

    \item We compare our reference analysis, fully accounting for the size of the separation bins of the 3PCF estimator, to a more efficient evaluation that neglects binning effects, finding large systematic errors on the bias parameters. Although such errors can be reduced by using
    a smaller bin size than the one adopted here ($10 \hMpc$), a careful assessment has to be made before adopting this approximation, also considering that a smaller bin implies a much larger covariance matrix, with all its significant numerical requirements.
    
    \item Finally, we consider relations among the bias parameters in order to reduce the parameter space. We confirmed that the simple local Lagrangian expression for the tidal bias leads to significant systematic errors in the 3PCF case as well. The relation between $b_2$ and $b_1$ obtained by \cite{Lazeyras2016} as a fit to numerical simulations, instead leads to results consistent with our reference case where all three parameters are left free.

\end{itemize}

With this work we have provided a solid test of 3-point correlation function as a useful tool in the determination of second-order bias parameters, showing constraints consistent to those from a bispectrum analysis. A fully numerical estimates of the 3PCF covariance matrix, however, is always a very numerically demanding task. We discussed here possible strategies to tackle this problem and provide ideal tests based on periodic-box simulations. We notice that we could consider other techniques to compress the information, avoiding the explicit need of a large set of mocks to analyse \citep[see e.g.][]{Slepian2015, Gualdi2020, Philcox2021, Fumagalli2022}. We will explore these techniques in more details and more realistic scenarios in future analyses, quantifying the effect of specific survey geometry, as well as test a full likelihood analysis in redshift space. Other possible extensions of this work go in the direction of testing more sophisticated models (e.g. \cite{Eggmeier2019}) to extract as much information as possible on both bias and cosmological parameters. 

%==================================================
\acknowledgments

We thank the anonymous referee for the comments and suggestions that helped improve the paper. We want to thank Luigi Guzzo, Elena Sarpa and Massimo Guidi for useful discussions,
Claudio Dalla Vecchia 
 for performing and sharing the  {\Min} simulations used in this work. The {\Min} simulations have been performed on the Hydra and Euclid clusters at the Max Planck Computing and Data Facility (MPCDF) in Garching.
The {\Pin} mocks were run on the GALILEO cluster at CINECA, thanks to an agreement with the University of Trieste. 
AV, EB and MM are partially supported by ASI/INAF agreement n. 2018-23-HH.0 ``Scientific activity for Euclid mission, Phase D".
EB, AV, PM and ES are also supported by INFN project ``InDark". 
EB and AV are also supported by MIUR/PRIN 2017 ``From Darklight to Dark Matter: understanding the galaxy-matter connection to measure the Universe".
EB is also supported by ASI/INAF agreement  n. 2017-14-H.O ``Unveiling Dark Matter and Missing Baryons in the high-energy sky". 
ES and PM are also partially supported by PRIN MIUR 2015 "Cosmology and Fundamental Physics: illuminating the Dark Universe with Euclid."  
MM acknowledges support from MIUR, PRIN 2017 (grant 20179ZF5KS).

%==================================================
%\paragraph{Note added.} This is also a good position for notes added after the paper has been written.

\setlength{\bibsep}{2pt plus 0.5ex}
\bibliographystyle{JHEP}
\nocite{*}
\bibliography{bibliography}

\end{document}